\documentclass[%
 reprint,
nofootinbib,
 amsmath,amssymb,
 aps,
]{revtex4-1}

\usepackage{graphicx}
\usepackage{dcolumn}
\usepackage{bm}
\usepackage[colorlinks=true]{hyperref}

\begin{document}

\title{The method of generating functions in exact scalar field inflationary cosmology}

\author{Sergey V. Chervon}
\email{chervon.sergey@gmail.com}
\affiliation{Astrophysics and Cosmology Research Unit, School of Mathematics, Statistics and Computer Science,
University of KwaZulu-Natal, Private Bag X54 001, Durban 4000, South Africa}
\affiliation{Bauman Moscow State Technical University, 2-nd Baumanskaya street, 5, Moscow, 105005, Russia}
\author{Igor V. Fomin}
\email{ingvor@inbox.ru}
\affiliation{Bauman Moscow State Technical University, 2-nd Baumanskaya street, 5, Moscow, 105005, Russia}
\author{Aroonkumar Beesham}
\email{beeshama@unizulu.ac.za}
\affiliation{Department of Mathematical Sciences, University of Zululand, Private Bag X1001, Kwa-Dlangezwa 3886, South Africa
}
\date{\today}

\begin{abstract}
The construction of exact solutions in scalar field inflationary cosmology is of growing interest.
In this work, we review the results which have been obtained with the help of one of the most effective methods, viz., the method of generating functions for the construction of exact solutions  in scalar field cosmology. We also include in the debate the superpotential method, which may be considered as the bridge to the slow roll approximation equations. Based on the review, we suggest a classification for the generating functions, and find a connection for all of them with the superpotential.
\begin{description}
\item[PACS numbers] 98.80.Bp, 98.80.Cq, 98.80.Es, 98.80.Jk
\end{description}
\end{abstract}

\maketitle
\affiliation{Ulyanovsk State Pedagogical University, Ulyanovsk, 100-years V.I. Lenin's Birthday Square, B.4, 432071, Russia\\
 Bauman Moskow State Technical University, 2-nd Baumanskaya street, 5, Moscow, 105005, Russia\\
University of Zululand, Richards Bay\\
Astrophysics and Cosmology Research Unit, School of Mathematics, Statistics and Computer Science,
University of KwaZulu-Natal, Private Bag X54 001, Durban 4000, South Africa }

\section{Early Inflation and the implemented scalar field }

Inflationary expansion of the universe during very early times, once the universe emerged from the quantum gravity (Planck) era, has been proposed  in the late 1970's and, mainly  in the beginning of the 1980's and is becoming more accepted as a necessary stage of the standard Big Bang theory model.  In the work of Starobinsky (1978) \cite{Starobinsky:1978}, nonsingular isotropic cosmological models with a massive scalar field were investigated. Although this work was concerned more with a bouncing cosmology rather than inflation, the field equations and the corresponding slow-roll solutions were already derived in this paper. These were later employed in the chaotic inflation model of Linde\cite{Linde:1983gd}.
The works by Starobinsky (1980)\cite{Starobinsky:1980te}, Guth (1981)\cite{Guth:1980zm} Linde (1982)\cite{Linde:1981mu}, Albrecht and Steinhardt (1982)\cite{Albrecht:1982wi} include the physical mechanism based on quantum corrections and phase transitions during the very early stage of the universe. Exponential (de Sitter) expansion is the feature of inflationary models which helped to solve the long standing problems of the standard Big Bang theory model: the horizon, flatness, homogeneity, isotropy and some other problems.

 In Sato's work (1981) \cite{Sato:1981ds}, the first-order phase transition model of the early Universe that leads to an exponential expansion which stretches domains much greater than the horizon scales was considered. Also, in  \cite{Sato:1981f} it was pointed out that fluctuations associated with the phase transition are exponentially stretched and then may play the role of seed fluctuations for large-scale structures. Further, the monopole problem was also shown to be resolved by exponential expansion by Einhorn and  Sato (1981) \cite{Einhorn:1980ik}. The chaotic inflation scenario proposed by Linde (1983) \cite{Linde:1983gd} differs from other previous versions since it is not based on the theory of high-temperature phase transition in the very early universe, but contains the locally homogeneous scalar field which is slowly rolling down to the minimum of the scalar field potential.

After that proposal, many investigations took place of the inflationary universe connected with a self-interacting scalar field as the source of gravitation in the Friedmann world. Let us briefly mention some interesting works concerning the study of a scalar field in inflationary cosmology.

Homogeneous isotropic cosmological models with a massive scalar field have been studied in the works \cite{Belinsky:1985zd,Piran:1985pl}. It was shown that inflationary stages are a fairly general property of most solutions in the considered model. The general conditions for inflation were investigated in the work \cite{Piran:1986dh}. It was found that under the lower limit for the amplitude of a scalar field, the universe naturally enters into and exits out of an inflationary phase. What is important is that such behavior takes place under a large variety of scalar potentials which are polynomial, logarithmic or exponential. It was also stated that a scalar field is essential for inflation \cite{Piran:1986dh}; it is unlikely that a vector or other non-scalar field will lead to inflation. The difference between scalar potentials in particle physics and those in cosmology has been stressed in the work \cite{Halliwell:1986ja}. The author wrote: "... we do not really know which theory of particle physics best describes the very early universe. One should therefore keep an open mind as to the form of $V(\phi)$." Halliwell chose the exponential potential and showed that it leads to a solution with power-law inflation, and that this solution is an attractor. Detailed investigations of power-law inflation have been carried out in the work \cite{Lucchin:1985yf}. The authors found the constraints on the model coming from the requirement of solving the horizon, flatness, reheating and perturbation-spectrum problems. It was stated also that these constraints can be suitably satisfied. An exact power-law inflationary solution possessing an exponential potential was given in the work \cite{Barrow:1987ia}.  The generic inhomogeneous generalization of this solution having both scalar and tensor superhorizon "hairs" was derived in the paper \cite{Muller:1989rp}.

Let us mention also the investigation carried out by Ivanov (1981) \cite{givanov81} where he found  exact solutions for a nonlinear scalar field in cosmology. The solutions he obtained included polynomial, trigonometric and exponential potentials. The method he used for searching for exact solutions  was subsequently called the Hamilton-Jacobi-like approach.

From the observational point of view, most results which can be related to observational data have been obtained from the so-called slow roll approximation of the cosmological dynamical equations \cite{Linde:1981mu}, \cite{Albrecht:1982wi}. Detailed investigations of various physical phenomena from particle physics and GUT theories for the period until the 1990's can be studied from the reviews \cite{Olive:1989nu,Olive:1990sb,Goldwirth:1991rj}. Our attention will be concentrated on exact solutions of inflationary models, the study of which started about ten years later, after inflationary cosmology had been proposed.

Thus we are going to present a brief review of the construction of exact solutions in the inflationary universe, i.e., the solutions of self-consistent Einstein and scalar field equations in Friedmann cosmology. The direct connection between scalar field cosmology and cosmology based on the perfect fluid stress-energy tensor needs to be mentioned. This connection is always valid except in the case of dust matter. Therefore we included the case of exact solutions for perfect fluid as the source of gravitation.

The construction of exact solutions in inflationary cosmology started with the work by Muslimov (1990) \cite{M90}. The results presented in that article will be discussed in Sec. \ref{Sect1.3}. Here we would like to mention that the very method and many interesting exact solutions presented in \cite{givanov81} have been reproduced and generalized in \cite{M90}. New methods and new sets of exact solutions have been developed in the work \cite{M90} as well.

Barrow \cite{Barrow:1990vx}  found a simple way to solve exactly the cosmological dynamic equations in terms of a pressure-density relationship. In this way he obtained the known power-law and de Sitter forms of inflation and new classes of behavior in which
the expansion scale factor increases as the exponent of some power of the cosmic time coordinate. The double-exponential law solution was obtained as well.

The work by Ellis and Madsen (1991) \cite{Ellis:1990wsa} was the first where "the inverse problem" was considered in the framework of cosmology. Usually one suggests that we know the scalar potential in the very early universe from HEP,  and our task is to find the scale factor and the scalar field as functions of  time. However Ellis and Madsen (1991)  \cite{Ellis:1990wsa} suggested starting from the given scale factor! Indeed, it is clear that the scale factor may be found from observational data. Then we may take into account this fact to find the potential and scalar field from the cosmological equations. This work was done and examples of exact solutions have been presented for the pure scalar field (without taking into account radiation which is also considered there).
Further this approach was developed in the works \cite{Chervon:1996ju, Chervon:1997yz}.

Reconstruction of models with a scalar field (quintessence) and dust-like particles (baryons and dark matter) from observational data was further developed by Starobinsky (1998) \cite{Starobinsky:1998fr}, Huterer and Turner (1999) \cite{Huterer:1998qv}, Nakamura and Chiba (1999) \cite{Nakamura:1998mt} using the luminosity data and in Starobinsky \cite{Starobinsky:1998fr} from the growth factor of inhomogeneities.

Our paper is organised as follows: In section II, we present the basic equations of scalar field cosmology. In section III, we discuss generating functions for finding solutions, and sections IV is devoted to the classification of the generating functions. The superpotential method is presented in section V and we conclude with section VI.



\section{Basic equations of scalar field cosmology}\label{Sec1.2}

We consider the model of a self-gravitating scalar field $\phi$ with the potential of self-interaction $V(\phi)$. The action of such a model is
\begin{equation}\label{1.1}
S=\int d^4 x \sqrt{-g}\left( \frac{R+\Lambda}{2\kappa}-\frac{1}{2}\phi_{,\mu} \phi_{,\nu}g^{\mu\nu} -V(\phi)\right),
\end{equation}
where $R$ is the curvature scalar, $\phi$ the scalar field, $\phi_\mu =\partial_\mu \phi $ the short representation of the partial derivative $ d\phi/dx^\mu$, $\kappa$ is Einstein's gravitational constant, and $\Lambda$ is the cosmological constant, which will mainly be included in the scalar field potential $V(\phi)$ as the constant part of it.

In the standard way one can obtain the energy-momentum tensor (EMT)

\begin{equation}\label{1.2}
T_{\mu\nu}^{(sf)}=\phi_{,\mu} \phi_{,\nu} - g_{\mu\nu}\left(\frac{1}{2} \phi_{,\rho}\phi^{,\rho}+V(\phi)\right),
\end{equation}
and the Einstein equation
\begin{equation}\label{1.3}
G_{\mu\nu}\equiv R_{\mu\nu}-\frac{1}{2} g_{\mu\nu}R= \kappa T_{\mu\nu}^{(sf)},
\end{equation}
may be represented through the trace of the EMT in the form

\begin{equation}\label{1.4}
R_{\mu\nu}= \kappa (-T_{\mu\nu}+\frac{1}{2} g_{\mu\nu}T)
= \phi_{,\mu} \phi_{,\nu}+g_{\mu\nu} V(\phi).
\end{equation}

Varying the action (\ref{1.1}) with the scalar field $\phi$, we obtain the dynamic equation of the scalar field
\begin{equation}\label{1.5}
-\nabla_\mu \nabla^\mu \phi + V'(\phi)=0,~~V'\equiv \frac{dV}{d\phi}.
\end{equation}
We consider the homogeneous and isotropic Universe as the spacetime with the Friedmann-Robertson-Walker (FRW) metric

\begin{equation}\label{1.6}
ds^2= -dt^2+ a^2(t)\left( \frac{dr^2}{1-\epsilon r^2}+r^2 \left(d\theta^2+\sin^2\theta d\varphi^2\right) \right),
\end{equation}
where $ \epsilon =0,~\epsilon =1,~\epsilon=-1 $ for the spatially-flat, closed and open universe, respectively.

The Einstein Eq. (\ref{1.3}) and the equation of the scalar field dynamics (\ref{1.5}) in the FRW metric (\ref{1.6}) lead  to the system of equations

\begin{eqnarray}\label{1.7}
\frac{\ddot{a}}{a}+\frac{2\dot{a}^2}{a^2}+\frac{2\epsilon}{a^2}=\kappa V(\phi),\\
\label{1.8}
-\frac{3\ddot{a}}{a}=\kappa \left( \dot{\phi}^2 - V(\phi) \right),\\
\label{1.9}
\ddot{\phi}+3\frac{\dot{a}}{a}\dot{\phi}+V'(\phi)=0.
\end{eqnarray}
Eqs. (\ref{1.7}) and (\ref{1.8}) can, in an equivalent way, be replaced by thier sum and the linear combination 3(\ref{1.7})+(\ref{1.8}). Including the Hubble parameter  $H=\dot{a}/a$, the system (\ref{1.7})-(\ref{1.9}) can be rewritten in the form

\begin{eqnarray}\label{1.10}
H^2+\frac{\epsilon}{a^2}=\frac{\kappa}{3}\left( \frac{1}{2} \dot{\phi}^2 +
V(\phi)\right),\\
\label{1.11}
\dot{H}-\frac{\epsilon}{a^2}=- \frac{\kappa}{2} \dot{\phi}^2, \\
\label{1.12}
\ddot{\phi}+3H\dot{\phi}+V'(\phi)=0.
\end{eqnarray}
We will refer to the system (\ref{1.10})-(\ref{1.12}) as {\it the Scalar Cosmology Equations (SCEs)}\label{SCEs}.

The representation above, Eqs. (\ref{1.10})-(\ref{1.12}) has some advantages for the derivation of any of the three  equations
(\ref{1.10})-(\ref{1.12}) from the other  two,  and differential consequences of them.

Another representation of the SCEs was first proposed by G. Ivanov \cite{givanov81}. Suggesting the dependence of the Hubble parameter $H$ on the scalar field $\phi $, the transformation of the equations (\ref{1.10})-(\ref{1.12}) for the spatially-flat universe ($ \epsilon=0$) to the form, which was called later the Hamilton-Jacobi-like form, was made. Eq. (\ref{1.11}) is transformed to

\begin{equation}\label{1.13}
H'=-\frac{\kappa}{2} \dot{\phi}.
\end{equation}
Squaring the above equation and making the substitution
$\dot{\phi}^2/2$, expressed in term of
$H'^2$, and substituting into (\ref{1.10}), one can obtain

\begin{equation}\label{1.14}
\frac{2}{3\kappa}\left[\frac{dH}{d\phi}\right]^2-H^2=-\frac{\kappa}{3}V(\phi).
\end{equation}

It is worthwhile to note that this procedure and the very equation (\ref{1.14}) have been obtained by G. Ivanov in 1981 \cite{givanov81}. Unfortunately this result was published in limited editions (in Russian) and it was not familiar outside the USSR. Fortunately, in 1990, in the work of A. Muslimov \cite{M90}, the Ivanov result was reproduced (Muslimov referenced the Ivanov article), and  some of the solutions were generalized as well. In the same year, within the detailed investigation of long-wavelength metric fluctuations in inflationary models, D. Salopek and J. Bond \cite{SB90} obtained the "separated Hamilton-Jacobi equation that
also governs the semiclassical phase of the wave functional". The obtained equation contains a couple of  scalar fields and definitely can be applied to a single scalar field.
Therefore we suggest that Eq. (\ref{1.14}) in the cosmological context  should be called {\it the Ivanov-Salopek-Bond (ISB) equation}.

\section{Generating function for solving the Ivanov-Salopek-Bond equation}\label{Sect1.3}

The structure of Eq. (\ref{1.14}) prompts us to find the form of the potential which can give the exact solution.
Indeed, let the potential  $V(\phi)$ be of the form
\begin{equation}\label{1.15}
V(\phi)=-\frac{2}{3\kappa}F'^2 +\left(F(\phi)+F_*\right)^2,
\end{equation}
where $F=F(\phi)$  is a $ C^1$ function on $\phi$ and $F_*=const.$
We will call the function $F(\phi)$   {\it the generating function}.
Comparing  Eq.(\ref{1.14}) with (\ref{1.15}), it is easy to find the solution of (\ref{1.14})
\begin{equation}\label{1.16}
H(\phi)= \sqrt{\frac{\kappa}{3}}(F(\phi)+F_*).
\end{equation}

Thus, one can directly from (\ref{1.14}) obtain the potential if the Hubble parameter is given. And vice versa, if one sets the potential in the form (\ref{1.15}), then the solution of
(\ref{1.14}) will be defined by the Eq. (\ref{1.16}).

\subsection{Potential in polynomial form}

As an example, let us choose the generating function
$F(\phi)$ as the finite series on degrees of the field $\phi$
\begin{equation}
F(\phi)=\sum_{k=0}^{p}\lambda_k\phi^k +F_*.
\end{equation}
Under this circumstance, the potential $V(\phi)$ takes the following form
\begin{equation}
V(\phi)=-\frac{2}{3\kappa}\left[\sum_{k=0}^{p-1}(k+1)\lambda_{k+1}\phi^k\right]^2+
\left[\sum_{k=0}^{p}\lambda_k\phi^k+F_* \right]^2.
\end{equation}
Let us consider the simple case when $F_*=0,~k=0,~p=1$. Then the potential
becomes
\begin{equation}
V(\phi)= -\frac{2}{3\kappa}\lambda_1^2 +\lambda_0^2 +2\lambda_0\lambda_1 \phi+\lambda_1^2\phi^2.
\end{equation}
The generating function $F(\phi)$ and the Hubble parameter are
\begin{equation}
F(\phi)=\lambda_0+\lambda_1 \phi, ~~H(\phi)=\sqrt{\frac{\kappa}{3}}(\lambda_0+\lambda_1 \phi).
\end{equation}

If we additionally set
$\lambda_0=0$, then we obtain the solution for the massive scalar field as in \cite{givanov81},
with $ \lambda_1^2=m^2/2 $  (for the sake of simplicity we also set $c=\hbar=1$). Thus the potential takes the form
\begin{equation}\label{1.21}
V(\phi)=\frac{m^2 \phi^2}{2}- \frac{m^2}{3\kappa}.
\end{equation}
Solving Eq.(\ref{1.13}) one can obtain the evolution of the scalar field
\begin{equation}\label{1.22}
\phi (t) = -m\sqrt{\dfrac{2}{3\kappa}}t+\phi_s= -m\sqrt{\dfrac{2}{3\kappa}}(t-t_*),~~\phi_s=m\sqrt{\dfrac{2}{3\kappa}}t_*.
\end{equation}
The index "s" ("singularity") here is related to the values at the initial time
  $t=0$, i.e., for a singularity in accordance with big bang theory.
The Hubble parameter
\begin{equation}
H= m \sqrt{\frac{\kappa}{6}}\phi,
\end{equation}
has a dependence $\phi$ on time
(\ref{1.22}), an this gives us a possibility to perform integration and obtain
the dependence of the scale factor on time
\begin{equation}\label{1.24}
a=a_s \exp \left(-\frac{m^2}{6}t^2+m\sqrt{\frac{\kappa}{6}}\phi_s t\right).
\end{equation}

Thus we obtained the exact solution for the potential
(\ref{1.21}), which is represented by the dependence of $\phi$ on time
(\ref{1.22}), and the scale factor on time
(\ref{1.24}).

Also, for the first time, solution (\ref{1.24}) was obtained in \cite{Linde:1983gd} by using the slow-roll approximation for
the potential $V(\phi)=m^{2}\phi^{2}/2$ in contrast to the shifted potential  (\ref{1.21}).

It is interesting to note that the same solution and its application for the calculation of the number of e-folds
 and scalar spectral parameter were found and developed later by Wang \cite{Wang:2001jx}.

When $\lambda_0 \neq 0$ the solution for the scale factor will differ by the factor
$a_s$ in front of the exponent
\begin{equation}\nonumber
a=a_s \exp \left(-\frac{m^2}{6}t^2+m\sqrt{\frac{\kappa}{6}}\phi_s t+\lambda_0\sqrt{\frac{\kappa}{3}}\right)=
\end{equation}
\begin{equation}\label{1.25}
\tilde{a}_s\exp \left(-\frac{m^2}{6}t^2+m\sqrt{\frac{\kappa}{6}}\phi_s t \right).
\end{equation}
Here $ \tilde{a}_s= a_s \exp \left(\lambda_0\sqrt{\kappa/3}\right)$.
The potential $V(\phi)$ then takes the form
\begin{equation}
V(\phi)=-\frac{2}{3\kappa}\lambda_1^2+\lambda_0^2 +2\lambda_0\lambda_1\phi +
\lambda_1^2\phi^2=\left(\frac{m\phi}{\sqrt{2}}+\lambda_0\right)^2-\frac{m^2}{3\kappa}.
\end{equation}

Let us note that the linear transformation of the field
without changing of the mass
\begin{equation}
\tilde{\phi}=\frac{\phi}{\sqrt{2}}+\frac{\lambda_0}{m}
\end{equation}
leads to the potential
(\ref{1.21}) for the field $ \tilde{\phi}$.

Let us consider the case when $k=1,2$.
The generating function $F(\phi)$ takes the form
\begin{equation}
F(\phi)=\lambda_1\phi+\lambda_2\phi^2.
\end{equation}
The potential $V(\phi)$ is
\begin{equation}
V(\phi)=-\frac{2}{3\kappa}(\lambda_1+2\lambda_2\phi)^2 +(\lambda_1\phi+\lambda_2\phi^2)^2.
\end{equation}
We can make a simplification by considering $\lambda_1=0$. The potential then takes the Higgs form
\begin{equation}
V(\phi)=-\frac{2}{3\kappa}(2\lambda_2\phi)^2 +(\lambda_2\phi^2)^2.
\end{equation}
Using the relation (\ref{1.16}), we may find $H$ and $H'$ expressed through $\phi$
\begin{equation}\label{1.31}
H=\sqrt{\frac{\kappa}{3}}\lambda_2 \phi^2,~~H'=2\sqrt{\frac{\kappa}{3}}\phi.
\end{equation}
Eq. (\ref{1.13}) takes the form
\begin{equation}
2\sqrt{\frac{\kappa}{3}}\phi=-\frac{\kappa}{2}\dot{\phi}.
\end{equation}
Performing the integration we find the dependence of
$\phi$ on time
\begin{equation}
\phi=\exp \left[-\frac{4}{\sqrt{3\kappa}}\lambda_2(t-t_*)\right].
\end{equation}
Substituting this result into
(\ref{1.31}) and performing the integration, we will find the scale factor
$a(t)$
\begin{equation}
a(t)=a_s \exp \left[-\frac{\lambda_2\kappa}{8}\exp \left(-\frac{8\lambda_2}{\sqrt{3\kappa}}(t-t_*)\right)\right].
\end{equation}
This is the double-exponental law solution \cite{givanov81}, \cite{Barrow:1990vx}.

To make a comparison with Ivanov's results (with subscript "I" in his notation) \cite{givanov81}, let us display the relations between the parameters of the model
$$
\mu_I=-\frac{16}{3\kappa}\lambda_2^2,~~\lambda_I=-4\lambda_2^2,~~ 3\kappa\mu_I=4\lambda_I.
$$
The case when $\lambda_1\neq 0$ leads to the scalar field
\begin{equation}
\phi=\frac{1}{2\lambda_2}\exp \left[-\frac{4}{\sqrt{3\kappa}}\lambda_2(t-t_*)-\frac{\lambda_1}{4\lambda_2}\right].
\end{equation}
The scale factor then takes the following form
\begin{equation}\label{1.36}
a(t)= a_s \exp \left[-\sqrt{\frac{\kappa}{3}}\frac{\lambda_1^2}{4\lambda_2}t-
\frac{\sqrt{3\kappa}}{8\lambda_2}\exp \left(-\frac{8\lambda_2}{\sqrt{3\kappa}}(t-t_*)\right)\right].
\end{equation}
It is useful to note the role of the addition of the constant $F_*$ to the function
$F(\phi)$.
In our presentation for $H(\phi)$ (\ref{1.16}), we can extract the constant part of the Hubble parameter
\begin{equation}
H(\phi)=\sqrt{\frac{\kappa}{3}}(F+F_*)= \tilde{H}(\phi)+H_*.
\end{equation}
The presence of the constant $H_*$ will be exhibited as the additional factor
for $a(t)$
\begin{equation}
a(t)=e^{H_*t}\exp \left( \int H dt\right).
\end{equation}

In the considered examples above (\ref{1.25}) and (\ref{1.36}),
such factors can be extracted explicitly.

\subsection{Trigonometric potential}

The solution for the potential which leads to the Sine-Gordon type equation was obtained in \cite{givanov81}.
Such a setting used the special choice of the additional parameter.
Let us consider this point in detail.

We choose the corresponding generating function $F(\phi)$  as
\begin{equation}
F(\phi)=A \sin (\lambda\phi), ~~A,~\lambda - const.
\end{equation}
Then the potential is

\begin{equation}
V(\phi)=-\frac{2A^2\lambda^2}{3\kappa}\cos^2 (\lambda\phi)+
A^2 \sin^2 (\lambda\phi).
\end{equation}
To obtain the potential suggested in \cite{givanov81}, it is enough to choose the parameter
$ \lambda$ in the following way: $ \lambda^2=\frac{3\kappa}{2}$. Such a choice leads to the potential
\begin{equation}
V(\phi)=-A^2 \cos \left(\sqrt{6\kappa}\phi \right).
\end{equation}
Equating the parameter $A^2=\mu$, we find the correspondence of the potential function with that presented in \cite{givanov81}.

The Hubble parameter in terms of the scalar field can be defined from
(\ref{1.16})
\begin{equation}\label{H_sin}
H(\phi)=A\sqrt{\frac{\kappa}{3}}\sin (\lambda\phi).
\end{equation}
Integrating equation (\ref{1.13}), it is necessary to consider the integral
$$
\int\frac{dx}{\cos x}
$$
which has various functional representations
\begin{eqnarray}
\nonumber
\int\frac{dx}{\cos x}=\ln \left|\tan \left( \frac{\pi}{4}+\frac{x}{2}\right)\right| +c_1=\\
 \frac{1}{2} \ln \frac{1+\sin x}{1-\sin x}+c_2, ~~ c_1,c_2 - const.
\end{eqnarray}
In \cite{givanov81} the first representation was shown.
We take the second representation in which the scalar field is defined from the relation
\begin{equation}
\sin (\lambda\phi)= \tanh \left(A \lambda \sqrt{2}(t-t_*)\right).
\end{equation}
Applying this result to (\ref{H_sin}) and performing the integration over time
$t$,  we obtain the scale factor
\begin{eqnarray}
\nonumber
a(t)=a_s \left[\cosh (A \lambda \sqrt{2}(t-t_*))\right]^{1/3}=\\
a_s \left[\cosh (A \sqrt{3\kappa}(t-t_*))\right]^{1/3}.
\end{eqnarray}

\subsection{Exponential potential}\label{exp_pwl}

The exponential potential in \cite{givanov81} is given as
\begin{equation}\label{1.46}
V(\varphi)=\alpha \exp (\beta\varphi),~~\alpha,~\beta - const.
\end{equation}
If we set
\begin{equation}\label{1.47}
F(\phi)=A \exp (\mu\phi),
\end{equation}
then the potential takes the form
\begin{equation}
V(\phi)= A^2 \left(1-\frac{2\mu^2}{3\kappa}\right)\exp (2 \mu\phi).
\end{equation}
Comparing this result to the original potential
(\ref{1.46}), we can find the relations
\begin{equation}\label{1.49}
\alpha=A^2\left(1-\frac{2\mu^2}{3\kappa}\right),~~\beta=2\mu,~~A=
\sqrt{\frac{\alpha}{1-\frac{\beta^2}{6\kappa}}}
\end{equation}

In accordance with the general procedure explained at the beginning of Sect. \ref{Sect1.3},  one can obtain
\begin{equation}
H(\phi)=\sqrt{\frac{\kappa}{3}}A \exp (\mu \phi).
\end{equation}
Then the dependence of the scalar field on time
$t$ has a logarithmic character
\begin{equation}
\phi (t)=-\frac{1}{\mu}\ln\left( \frac{2A\mu^2}{\sqrt{3\kappa}}t \right)+\phi_s.
\end{equation}
The scale factor is evaluated via a power law
\begin{equation}\label{1.52}
a(t)=a_s (t-t_*)^{\kappa/2\mu^2}.
\end{equation}
An addition of the constant $F_*$ to $F(\phi)$ leads to the generalization of the solution
(\ref{1.52})
\begin{equation}
a(t)=a_s e^{H_*t}(t-t_*)^{\kappa/2\mu^2},~~H_*=\sqrt{\frac{\kappa}{3}}F_*.
\end{equation}
This is the exponential power law solution.
Then the potential acquires the additional terms
\begin{equation}\label{1.54}
V(\phi)= A^2 \left(1-\frac{2\mu^2}{3\kappa}\right)\exp (2 \mu\phi)+
2AF_*e^{\mu\phi}+F_*^2.
\end{equation}

Muslimov (1990) \cite{M90} found the generalization of Ivanov's solution for the exponential potential.
Let us represent this, which contains both solutions.

If we take the generating function $F(\phi)$  in the form (\ref{1.47}) with the potential (\ref{1.54})
then
\begin{equation}
H(\phi)=\sqrt{{\frac{\kappa}{3}}} \left( A e^{\mu\phi}+F_* \right).
\end{equation}
We can find that
\begin{equation}
H'=\frac{\kappa}{3} A\mu e^{\mu\phi}.
\end{equation}
Integrating (\ref{1.13}), we obtain
\begin{equation}
e^{\mu\phi}= \left(\frac{2A\mu^2}{\sqrt{3\kappa}}t + v_*\right)^{-1},
\end{equation}
where  $v_*$ is a constant of integration. Finally we find
\begin{equation}\label{1.58}
\phi (t)=-\frac{1}{\mu}\ln \left(\frac{2A\mu^2}{\sqrt{3\kappa}}t + v_*\right).
\end{equation}
To obtain Ivanov's solution \cite{givanov81}, we set
$v_*=0$ and take into account the relations
(\ref{1.49}).

To obtain Muslimov's solution \cite{M90}, we set
$v_*=1$ and take into account the relations below
$$
F_*=0,~~\Lambda=A^2 \left(1-\frac{2\mu^2}{3\kappa}\right),~~ A=2/\mu,~~B=A/\sqrt{3}
$$

The solution (\ref{1.58}) without restrictions on the parameter $v_*$ gives some generalization.

\subsection{The solution with an inverse potential}

The potential in \cite{M90} was presented in the following way
\begin{equation}
V(\phi)=m^2 \phi^{-\beta}\left(1-\frac{1}{6}\beta^2 \phi^{-2}\right),~~\beta>0.
\end{equation}
It is not difficult to check that the same potential can be obtained from the generating function
\begin{equation}
F(\phi)=m\phi^{-\beta/2}.
\end{equation}
Therefore it is clear that the solution can be obtained by a general scheme.
The Hubble parameter is
\begin{equation}
H(\phi)=\sqrt{\frac{\kappa}{3}}m \phi^{-\beta/2}+H_*
\end{equation}
As we know the influence of $H_*$ on the result,  we may take
$H_*=0$ for the sake of simplicity.

Integrating (\ref{1.13}) we can find the dependence of the scalar field on time
\begin{eqnarray}
\phi(t)=\left[ K_1(t-t_*)\right]^{2/(\beta+4)} +\phi_*, \\
\nonumber
K_1=\sqrt{\frac{\kappa}{3}}\left(\frac{\beta+4}{2m\beta}\right).
\end{eqnarray}
This result leads to the time dependence of the Hubble parameter
\begin{equation}
H(t)=\sqrt{\frac{\kappa}{3}}m\left[K_1 (t-t_*)\right]^{-\beta/(\beta+4)}.
\end{equation}
Then the scale factor is

\begin{equation}
a=a_s m\sqrt{\frac{\kappa}{3}}\left(\frac{\beta+4}{4}\right) K_1^{-\beta/(\beta+4)}
\exp \left((t-t_*)^{4/(\beta+4)}\right).
\end{equation}
The solution of such a type can be confronted with both a very early and late time universe. Similar solutions have been obtained in \cite{Barrow:1990vx}.

\subsection{The solution with an intermediate (hyperbolic) function}

In the range of the results described above, Muslimov \cite{M90} suggested a new original approach for solving the scalar field cosmology equation.
To simplify calculations, let us, following \cite{M90}, introduce a new variable
\begin{equation}
x=\sqrt{\frac{3\kappa}{2}}\phi
\end{equation}
and the potential function
\begin{equation}
f^2=\frac{\kappa}{3}\left|V(\phi)\right|.
\end{equation}

For this notation, the Ivanov-Salopek-Bond (ISB) equation
(\ref{1.14}) reduces to
\begin{equation}
\left(H'_x\right)^2-H^2=\mp f^2
\end{equation}
The upper case corresponds to the positive sign of the potential.
It is interesting to mention that an equation of this type was studied by Mitrinovitch in 1937 \cite{Mitrinovitch:1937ds}.

Let us search for a solution in the form %
\begin{equation}
H(x)=f(x)\cosh \left(\coth^{-1}y(x)\right),~~y>1,
\end{equation}
 The other choice is for the lower sign
\begin{equation}
H(x)=f(x)\sinh \left(\tanh^{-1}y(x)\right),~~y<1.
\end{equation}
Here we included inverse hyperbolical tangents instead of inverse hyperbolical cotangents as in Muslimov's work with the aim of avoiding plus-minus signs in the final equation.
Using the formulae above, for transition to the function
$y(x)$, we obtain

\begin{equation}\label{1.70}
\left[f'\cosh u+f \sinh u ~ u'\right]^2-\left[f(x)\cosh u\right]^2=-f^2.
\end{equation}
Here, for the sake of briefness, we introduce the function
$u$  in the following way
\begin{equation}\label{1.71}
u(y)=\coth^{-1}y(x).
\end{equation}

We can shift the second term on the left hand side of (\ref{1.70}) to the right hand side and, using the property of the hyperbolic function, take the root on the left and right hand side of the equation (considering all values as positive). As the result, we obtain the equation
\begin{equation}
f' \cosh u +f \sinh u u'=f \sinh u.
\end{equation}

The transition to the function $y(x)$ is performed by inverse substitution of
(\ref{1.71}) and by introducing the derivative
\begin{equation}
u'=\frac{y'}{1-y^2}.
\end{equation}
Finally we arrive at the following relation
\begin{equation}
\frac{f'}{f}=\tanh (\coth^{-1}y)\left(1-\frac{y'}{1-y^2}\right).
\end{equation}
After simple algebraic transformations, we acquire the Abel equation
\begin{equation}\label{1.75}
y'=\frac{f'}{f}y^3-y^2-\frac{f'}{f}y+1.
\end{equation}

Repeating the same procedure for the lower case, we once again arrive at the equation
(\ref{1.75}).

\subsection{Kim's exact solutions}
H.-C.Kim in the article \cite{Kim:2012dn} proposed the generating function $G(\phi)$, which we can compare with the Ivanov
generating function $F(\phi)$  (\ref{1.15})-(\ref{1.16}) with $F_{\ast}=0$.
The Hubble parameter  in the work \cite{Kim:2012dn} is represented as
\begin{equation}\label{76}
H(\phi,\dot \phi) = -\frac{1}{3 \dot \phi}\frac{dF^2(\phi)}{d\phi} .
\end{equation}

To obtain the relation between the generating functions  $F(\phi)$ and $G(\phi)$
we may simply equalise
$F(\phi)\equiv G(\phi)$.
With the representation (\ref{76}) the solution of the SCEs can be obtained with the following formulae
\begin{eqnarray}
V(\phi) = F^2(\phi) -\frac23[F'(\phi)]^2,
\end{eqnarray}
\begin{equation}
\dot \phi =  -\frac{2}{\sqrt{3}} F'(\phi), \qquad H = \frac{\dot a}{a} = \frac{1}{\sqrt3}F(\phi) .
\end{equation}

In the work \cite{Kim:2012dn},  two generating functions were considered.
First of them is for the constant potential $V(\phi) = \Lambda >0$ :
\begin{eqnarray}
F(\phi) = \sqrt{\Lambda}.
\end{eqnarray}
This gives De Sitter space-time with the scale factor expanding exponentially.
\begin{eqnarray}
\dot \phi = 0, \;     H =H_I\equiv \sqrt{\Lambda/3}.
\end{eqnarray}

The second generating function is given by
\begin{eqnarray}
F(\phi) = \frac{e^{\sqrt{\frac32}\phi} +\Lambda e^{-\sqrt{\frac32}\phi }}{2}.
\end{eqnarray}
The scalar field and the scale factor behave as
\begin{eqnarray}
\phi = \sqrt{\frac23}\log\left(\sqrt{3}H_I \tanh\big(\frac{3H_I}2t\big)\right), \\
a(t) =a_0 \sinh^{1/3} (3H_I t).
\end{eqnarray}

The third generating function 
\begin{equation}
F(\phi) = \sqrt{\Lambda}\left(1+ \frac{\mu}{n+1} |\phi|^{n+1}\right)
\end{equation}
leads to the power-like potential
\begin{equation}
V(\phi) = \Lambda\left(1+ \frac{\mu}{n+1} |\phi|^{n+1}\right)^2- \frac23 \Lambda \mu^2\phi^{2n},
\end{equation}
where $ \mu, n$ are constants.
The scalar field and the scale factor are
\begin{eqnarray}
\phi(t) &=& \left(2(n-1)\mu H_I t\right)^{-\frac1{n-1}},\nonumber \\
a(t) &=& a_0\exp\left(H_I t-\frac{\left(2(n-1)\mu
            H_I t\right)^{-\frac{2}{n-1}}}{2(n+1)}\right),
\end{eqnarray}
where we set $\phi_{0}=\phi(0)=0$ at the time $t=t_{0}$.

\subsection{Exact solutions for constant-roll inflation}

Now, we consider inflationary models with the additional condition
\begin{equation}
\label{constant-roll}
\ddot{\phi}=-(3+\alpha)H\dot{\phi},
\end{equation}
where $\alpha$ is arbitrary constant parameter.
Standard slow-roll inflation occurs when $\alpha\simeq-3$ while for the other ultra-slow-roll regime, one has $\alpha=0$.
Such models correspond to the constant-roll inflation models which interpolate between these two regimes, and were considered in the papers \cite{Martin:2012pe, Motohashi:2014ppa}. Also, this approach was used for analysis of cosmological models in  $f(R)$ gravity on the basis of conformal transformations from the Jordan frame to the Einstein frame which lead to Ivanov-Salopek-Bond (Hamilton-Jacobi-like) equations (\ref{1.13})--(\ref{1.14}) \cite{Motohashi:2017vdc}. In the works \cite{Motohashi:2017aob,Motohashi:2017vdc}, it was shown that these types of models satisfy the latest observational constraints.

Eq. (\ref{1.13}), in the system of units with $\kappa=1$, and Eq. (\ref{constant-roll})
gives the condition on the Hubble parameter \cite{Motohashi:2014ppa}
\begin{equation}
\frac{d^{2}H}{d\phi^{2}}=\frac{3+\alpha}{2}H
\end{equation}
with general solution
\begin{eqnarray}
\label{constant-roll_H}
\nonumber
H(\phi)=C_{1}\exp\left(\sqrt{\frac{3+\alpha}{2}}\phi\right)+\\
C_{2}\exp\left(-\sqrt{\frac{3+\alpha}{2}}\phi\right).
\end{eqnarray}
From Eq. (\ref{1.14}) one has the potential
\begin{eqnarray}
\label{constant-roll_V}
\nonumber
V(\phi)=-\alpha C^{2}_{1}\exp\left(\sqrt{2(3+\alpha)}\phi\right)+\\
2(6+\alpha)C_{1}C_{2}-\alpha C^{2}_{2}\exp\left(-\sqrt{2(3+\alpha)}\phi\right).
\end{eqnarray}
On the basis of the general solutions (\ref{constant-roll_H})--(\ref{constant-roll_V}), one can find
the specific ones \cite{Motohashi:2014ppa}:

$\bullet$ The solutions for $\alpha>-3$ only:

 Power-law inflation\footnote{In the original work \cite{Motohashi:2014ppa}, the time $t$ was omitted in the expression for the Hubble parameter  (\ref{constant-roll_H_powerlaw}), but the scale factor was correct.}, which was already considered in subsection (\ref{exp_pwl})
\begin{eqnarray}
&&H=M e^{\sqrt{\frac{3+\alpha}{2}}\phi},\\
&&V(\phi)=- \alpha M^2 \exp\left(\sqrt{2(3+\alpha)}\phi\right),\\
&&\phi=-\sqrt{\frac{2}{3+\alpha}}\ln[(3+\alpha)Mt], \\
&&H=\frac{1}{(3+\alpha)t}\label{constant-roll_H_powerlaw},\\
&&a\propto t^{\frac{1}{3+\alpha}},
\end{eqnarray}
where $M$ is an integration constant.

$\bullet$ The solutions for both cases $\alpha>-3$ and $\alpha>-3$:

The inflation model with hyperbolic scale factor in the case of $\alpha>-3$:
\begin{eqnarray}
&&H=M \cosh\left(\sqrt{\frac{3+\alpha}{2}}\phi\right), \\
&&V(\phi)=3M^2\left(1+\frac{\alpha}{6}\left[1-\cosh(\sqrt{2(3+\alpha)}\phi)\right]\right),\\
&&\phi=\sqrt{\frac{2}{3+\alpha}}\ln\left[\coth\left(\frac{3+\alpha}{2}Mt\right)\right],\\
&&H=M \coth [ (3 + \alpha) M t], \\
&&a\propto \sinh^{1/(3+\alpha)}[(3+\alpha)Mt].
\end{eqnarray}
This solution is equivalent to a solution found in \cite{Barrow:1994nt} but in a different context.

The inflation model with hyperbolic scale factor in the case of $\alpha<-3$:
\begin{eqnarray}
&&H=M \cosh\left(\sqrt{\frac{3+\alpha}{2}}\phi\right), \\
&&V(\phi)=3M^2\left(1+\frac{\alpha}{6}\left[1-\cos(\sqrt{2(3+\alpha)}\phi)\right]\right),\\
&&\phi=2\sqrt{\frac{2}{|3+\alpha|}}{\rm arctan}(e^{|3+\alpha|Mt}),\\
&&H=-M \tanh [ (3 + \alpha) M t], \\
&&a\propto \cosh^{1/(3+\alpha)}[(3+\alpha)Mt].
\end{eqnarray}
This is the a particular case of hilltop inflation which is considered in paper \cite{Boubekeur:2005zm}.

The model with oscillating scale factor:
\begin{eqnarray}
&&H=M \sinh\left(\sqrt{\frac{3+\alpha}{2}}\phi\right), \\
&&V(\phi)=3M^2\left(1+\frac{\alpha}{6}\left[1-\cosh(\sqrt{2(3+\alpha)}\phi)\right]\right),\\
&&\phi=-2\sqrt{\frac{2}{3+\alpha}}{\rm arctanh}\left[\tan\left(\frac{3+\alpha}{2}Mt\right)\right],\\
&&H=-M \tan [ (3 + \alpha) M t], \\
&&a\propto \cos^{1/(3+\alpha)}[(3+\alpha)Mt].\label{constant-roll_a_cos}
\end{eqnarray}
From the scale factor (\ref{constant-roll_a_cos}) one has $\ddot a(t)<0$; nevertheless, this solution corresponds to
the oscillatory dynamics of the early universe.

\section{The classification of generating functions}

We propose a classification of generating functions as they appeared in the literature in chronological order.
As the first class, we name the Ivanov generating function $F(\phi)$  (\ref{1.15})-(\ref{1.16}). It was not represented in
direct form in \cite{givanov81}, but in the present publication for the first time, using this generating function, we recover and generalise all solutions of Ivanov's work \cite{givanov81}.

Now we continue the classification of the generating functions that occur in the literature.

\subsection{The second class of generating functions}
L. P. Chimento, A. E. Cossarini, and A. S. Jakubi (1993--1995)
in the papers \cite{Chimento:2012rx,Chimento:1995da} represented the potential of the scalar field in the form
\begin{equation}
V[\phi (a)] = {\frac{F(a)}{a^{6}}},
\end{equation}
where $F(a)$ is a new type of generating function.
The scalar field equation can be integrated 
and it yields 
\begin{equation}
\frac{1}{2}\dot{\phi}^{2}+V(\phi )-\frac{6}{a^6}\int da{\frac Fa}=
 \frac{C}{a^6},
\end{equation}
where $C$ is an arbitrary integration constant.
Thus, the problem of generation of exact solutions has reduced to the quadratures:
\begin{equation}
\label{12}
\Delta t={\sqrt{3}\int \frac{da}a}\left[ {\frac 6{a^6}\ \int }da{
\frac Fa+\frac C{a^6}}\right] ^{-1/2},
\end{equation}

\begin{equation}
\Delta \phi ={\sqrt{6}\int \frac{da}a\left[ \frac{-F+6\int daF/a+C
}{6\int daF/a+C}\right] }^{1/2},
\end{equation}
where $\Delta t\equiv t-t_0$, $\Delta \phi \equiv \phi -\phi _0$
and $t_0$, $\phi _0$ are arbitrary integration constants.

For the generating function
\begin{equation}
F(a)=B a^s\left( b+a^s\right) ^n,
\end{equation}
where $B(>0)$, $b(> 0)$, $s$ and $n$ are constants and $s(n+1)=6$.  If we take $C=0$, the potential is simplified to
\begin{equation}
V(\phi )=B\left[ \cosh \left( {\frac s{2\sqrt{6}}}\Delta \phi \right) \right]^{2n}.
\end{equation}
This potential has a non-vanishing minimum at $\Delta \phi =0$ for
$s>0$, which is equivalent to an effective cosmological constant.
When $s<0$ , the origin becomes a maximum, and the potential vanishes
exponentially for large $\phi $.

In  \cite{Chimento:2012rx} the equation (\ref{12}) was evaluated for some values of $s$:
\begin{equation}
\Delta t={\sqrt{\frac{3}{B} }}\left[ {\rm arcsinh} \left({\frac a{\sqrt{b}}}\right)
-{\frac a{(b+a^2)^{1/2}} }\right],  s=2.
\end{equation}

\begin{equation}
a=\left\{b\left[\exp \left(\sqrt{3B}\Delta
t\right)-1\right]\right\}^{1/3}, \qquad s=3.
\end{equation}
For $s>0$, the evolution begins from a singularity as
${\Delta t}^{1/3}$ and is asymptotically De~Sitter with
$\Delta\phi\rightarrow 0$ for $t\rightarrow\infty$. On the other hand, for
$s<0$ the evolution has a deflationary behaviour from a De Sitter era in the
far past to a Friedmann   behavior ${\Delta t}^{1/3}$ when
$t\rightarrow\infty$.

\subsection{The third class of generating functions}
In the paper  F. E. Schunck and E. W. Mielke (1994), \cite{Schunck:1994yd} the equations (\ref{1.7})-(\ref{1.9}) are written in the following form
\begin{equation}
\dot H =V(H) - 3H^2,
\end{equation}
\begin{equation}
\dot \phi =\pm \sqrt {2}\sqrt{3H^2 -V(H)},
\end{equation}
where $V(\phi ) = V(\phi (t)) = V(\phi (t(H))) = V(H)$.

The potential as a function of the Hubble parameter $V=V(H)$ is defined as
\begin{equation}
V(H)=3H^2+g(H).
\end{equation}
By  choice of the graceful exit function $g(H)$, the exact solutions of the SCES 
can be generated.

The choice for power--law and intermediate inflation are
\begin{equation}
g(H) =-AH^{n},
\end{equation}
where $n$ is real and $A$ is a positive constant.

For $n=0$, the following solutions were found:
\begin{equation}
H(t)=- (At +C_1),
\end{equation}
\begin{equation}
a(t) = a_0 \exp \left ( -\frac{1}{2A} (A t+C_1)^2 + C_2 \right),
\end{equation}
\begin{equation}
\phi (t) = \pm \sqrt{ \frac {2A}{\kappa } } (At+C_1-C_3),
\end{equation}
\begin{equation}
V(\phi ) =3\left (\sqrt{\frac{\phi}{2A}}+C_3\right )^2-A.
\end{equation}

For $n=1$
\begin{equation}
H(t)= C_1\exp (-A t),
\end{equation}
\begin{equation}
a(t) = a_0 \exp \left (-\frac{C_1}{A} \exp (-At)+\frac {C_2}{A}\right),
\end{equation}
\begin{equation}
\phi (t) =\pm \sqrt{\frac {8}{A}}\left [ \sqrt{C_1}\exp\left (\frac {-At}{2}-C_3 \right )\right],
\end{equation}
\begin{equation}
V(\phi ) =\frac {A}{8} e^{2 C_3}\phi^2\left (\frac{3A}{8} e^{2 C_3} \phi^2-A\right).
\end{equation}
For $n=2$
\begin{equation}
H(t)=\frac {1}{At+C_1},
\end{equation}
\begin{equation}
a(t) =a_0 (C_2 (At+C_1))^{1/A},
\end{equation}
\begin{equation}
\phi (t) = \pm \sqrt{\frac{2}{A}}\ln \left ( \frac {1}{C_3 (At+C_1)}\right),
\end{equation}
\begin{equation}
V(\phi ) =(3-A)C^2_3\exp (\pm \sqrt{2A}\phi),
\end{equation}
For $n\neq 0,1,2$
\begin{equation}
H =\left ( A (n-1) (t+C_1) \right )^{1/(1-n)},
\end{equation}
\begin{equation}
a(t) = a_0\exp \left [ (A (n-1))^{1/(1-n)} \frac {1-n}{2-n}(t+C_1)^{(2-n)/(1-n)} \right],
\end{equation}
\begin{equation}
\phi (t) + C_3 =\sqrt { \frac {2}{A}}\frac {2}{2-n} \Bigl [A (n-1) (t+C_1) \Bigr]^{(2-n)/(2(1-n))},
\end{equation}
\begin{gather}
V(\phi ) =\sqrt{\frac{A}{8}}(2-n)\left (\phi + C_3 \right )^{2/(2-n)}\times \nonumber \\
\left ( 3\frac{A}{8} (2-n)^2\left (\phi+C_3\right )^2
 - A\left (\frac {A}{8} \right )^{n/2} (2-n)^n
 \left (\phi+C_3 \right )^n\right).
\end{gather}

\subsection{The fourth class of generating functions}

 A. T. Kruger and J. W. Norbury (2000)  \cite{Kruger:2000nra}
proposed the generating function $F =F(\phi)$ via
\begin{equation}
\dot\phi \equiv \pm \sqrt{( F -1)V}.
\label{1-10}
\end{equation}
Then the scalar field dynamic equation transforms to
\begin{equation}
\ddot{\phi} \pm \sqrt{\frac{3}{2}}V\sqrt {F^2-1}+V^\prime  =0.
\label{1-11}
\end{equation}
Equations (\ref{1-10}) and
(\ref{1-11}) form a set of coupled equations which are equivalent to equation
(\ref{1.12}).  They can be uncoupled by differentiating (\ref{1-10}) with respect
to time and substituting
$\ddot{\phi}$ in (\ref{1-11}).
Using the evident relation $\ddot{\phi} = d (\dot \phi^2)/(2d t)$, by taking time derivative of the square of Eq. (\ref{1-10}), we have
\begin{equation}
\ddot{\phi}=\frac{1}{2}\left[(  F-1) V^\prime  +  V F^\prime \right],
\label{1-12}
\end{equation}
where $ F^\prime \equiv dF/d \phi$. Then,  inserting the expression for $\ddot{\phi}$ into
(\ref{1-11}) we obtain 
\begin{equation}
\label{1-13}
( F+1) V^\prime  + V F^\prime \pm \sqrt{6} V\sqrt {F^2-1} = 0.
\end{equation}
It can be seen that if one chooses $F \equiv F(\phi)$,
then equation (\ref{1-13}) is always separable and the potential is given
by
\begin{equation}
V =\beta \exp  \left( - \int   \frac{  F^\prime \pm \sqrt{6} \sqrt
{F^2-1} }{ F +1}d\phi \right),
\label{1-21-1}\
\end{equation}
where $F \equiv F(\phi)$ and $\beta$ is a constant.



The choice of the generating function in the work \cite{Kruger:2000nra},
is $F= \cosh(\lambda \phi)$. The the solution can be obtained from (\ref{1-21-1}) and (\ref{1-10}):
\begin{equation}
V(\phi) = C \left( 1 + \cosh \lambda \phi \right)^{\mp (2\sqrt{6}/\lambda) - 1},
\label{1-20ab}
\end{equation}
%
These equations are only consistent if all upper or all lower signs are taken, i.e., one should not mix
upper and lower signs. For the sake of simplicity, we chose $\lambda = \sqrt{6}$.

The upper $"-"$ sign in equation(\ref{1-20ab}), corresponding to the upper $"+"$ sign in the definition
(\ref{1-10}),  gives
\begin{equation}
V(\phi)=\frac{C}{(1+ \cosh \lambda \phi)^3}.
\label{1-111}
\end{equation}
The lower $"+"$ sign in (\ref{1-20ab}),  corresponding to the lower $"-"$ sign in the definition
(\ref{1-10}), leads to 
\begin{equation}
V(\phi)= C (1+ \cosh \lambda \phi).
\label{1-111a}
\end{equation}
Using this result (\ref{1-111a})
in equation (\ref{1-10}), we obtain
\begin{equation}
\dot{\phi}= - \sqrt {C} \sinh \lambda \phi.
\label{1-16}
\end{equation}
The solution for $\phi(t)$ yields
\begin{eqnarray}
\nonumber
\phi(t)=\frac{2}{\lambda} \coth^{-1} \left\{\exp \left[ \lambda  \sqrt {C} (t -
D)\right]\right\}
 =\\
 \frac{1}{\lambda}\ln \left( \frac{\exp \left[  \lambda \sqrt {C} (t -
D)\right] + 1}{\exp
\left[ \lambda
\sqrt { C} (t - D)\right] -1} \right),
\label{1-17}
\end{eqnarray}
where $D$ is a constant. The form of the potential suggests that
$\phi(t)$ be a function that decreases from an initial maximum value similar to the chaotic inflation
model.  We can choose $D=0$ and so (\ref{1-17})  can be written as
\begin{equation}
\phi(t)= \frac{1}{\lambda}\ln \left[\frac{\exp(\lambda\sqrt{C} t) +
1}{\exp( \lambda \sqrt{C} t) - 1}\right].
\label{1-18}
\end{equation}
This solution can be used to determine the evolution of the scale factor
with time for the expanding ($"+"$ square root) solution
\begin{equation}
a(t) = a_{0}\left[\exp(2 \lambda\sqrt {C}\hspace{1.2mm} t)-1\right]^{\frac{1}{3}}.
\label{1-22}
\end{equation}
\subsection{The fifth class of generating functions}

T. Charters and J. P. Mimoso (2010)
in the paper \cite{Charters:2009ku} proposed a generating function in the following form
\begin{equation}
\label{x}
x(\phi)=\dot{\phi}/H.
\end{equation}
The scalar field's dynamical equations are transformed to the following
\begin{equation}
\label{xV}
V(\phi)=A\left(3-\frac{1}{2}x^{2}(\phi)\right)\exp\left(-\int x(\phi)d\phi\right),
\end{equation}
\begin{equation}
\label{xH}
H(\phi)=\pm\sqrt{A}\exp\left(-\frac{1}{2}\int x(\phi)d\phi\right),
\end{equation}
\begin{equation}
\label{xphi}
\dot{\phi}^{2}=x^{2}(\phi)\exp\left(-\int x(\phi)d\phi\right).
\end{equation}
Further, in \cite{Charters:2009ku} the known and new exact solutions were obtained by the special choice of the generating function $x(\phi)$.

For example, the following type of $x(\phi)=\lambda\phi$ corresponds to the generalization of the Easther solution~\cite{Easther:1995pc}:
\begin{eqnarray}
    \label{eq:VEasther}
    V(\phi)=A\left(3-\lambda^2\phi^2/2\right)e^{\lambda\phi^2/2},
  \end{eqnarray}
\begin{eqnarray}
  a(\phi)&=& a_0 \phi^{1/\lambda} , \\
  t(\phi)&=&\frac{1}{2\lambda\sqrt{A}}\left[\mathrm{Ei}\left(\frac{\lambda\phi_0^2}{4}\right)-\mathrm{Ei}\left(\frac{\lambda\phi^2}{4}\right)\right],
\end{eqnarray}
where $\mathrm{Ei}$ is the exponential integral function.

\subsection{The sixth class of generating functions}
 T. Harko, F. S. N. Lobo, and M. K. Mak (2014) \cite{Harko:2013gha} introduced a new generating function $G$ with dynamical equations
 \begin{equation}
\frac{dG}{d\phi }+\frac{1}{2V}\frac{dV}{d\phi }\coth G+\sqrt{\frac{3}{2}}=0,
\label{fin}
\end{equation}
 \begin{equation}
\dot{H}=-\frac{1}{2}\dot{\phi}^{2}=-V\sinh^{2}G,
\label{fin1}
\end{equation}
where the function $G$ can be defined via the scalar field from the equation (\ref{fin1}) as
\begin{equation}  \label{G}
G(\phi)=\mathrm{arccosh} \sqrt{1+\frac{\dot{\phi}^{2}}{2V(\phi)}}.
\end{equation}
In \cite{Harko:2013gha} the authors considered the case when the scalar field
potential can be represented as the function of $G$ in the form
\begin{equation}
\frac{1}{2V}\frac{dV}{d\phi }=\sqrt{\frac{3}{2}}\;\alpha _{1}\,\tanh G,
\end{equation}%
where $\alpha _{1}$ is an arbitrary constant. With this choice, the
evolution equation takes the simple form
\begin{equation}
\frac{dG}{d\phi }=\sqrt{\frac{3}{2}}\left( 1+\alpha _{1}\right) .
\end{equation}%
Performing the integration one can obtain the general solution
\begin{equation}\label{138}
G\left( \phi \right) =\sqrt{\frac{3}{2}}\left( 1+\alpha _{1}\right) \left(
\phi -\phi _{0}\right) ,
\end{equation}%
where $\phi _{0}$ is an arbitrary constant of integration. Using
the form (\ref{138}) of $G$, one can obtain the self-interaction potential of
the scalar field and the scale factor
\begin{equation}
V\left( \phi \right) =V_{0}\cosh ^{\frac{2\alpha _{1}}{1+\alpha _{1}}}\left[
\sqrt{\frac{3}{2}}\left( 1+\alpha _{1}\right) \left( \phi -\phi _{0}\right) %
\right] ,  \label{kkk}
\end{equation}
\begin{equation}
a=a_{0}\sinh ^{\frac{1}{3\left( 1+\alpha _1\right) }}\left[ \sqrt{\frac{3}{2}
}\left( 1+\alpha _1\right) \left( \phi -\phi _{0}\right) \right].
\end{equation}

A simple solution of the gravitational field equations for a power-law type
scalar field potential can be obtained by assuming for the function $G$ the
following form
\begin{equation}
G=\mathrm{arccoth}\left( \sqrt{\frac{3}{2}}\frac{\phi }{\alpha _{2}}\right)
,\qquad \alpha _{2}=\mathrm{constant}.
\end{equation}

With this choice of $G$,  Eq.~(\ref{fin}) immediately provides the
scalar field potential given by
\begin{equation}
V\left( \phi \right) =V_{0}\left( \frac{\phi }{\alpha _{2}}\right)
^{-2\left( \alpha _{2}+1\right) }\left[ \frac{3}{2}\left( \frac{\phi }{%
\alpha _{2}}\right) ^{2}-1\right] ,  \label{mmm}
\end{equation}%
where $V_{0}$ is an arbitrary constant of integration. The time dependence
of the scalar field is given by a simple power law,
\begin{equation}
\frac{\phi (t)}{\alpha _{2}}=\left[ \frac{\sqrt{2V_{0}}\left( \alpha
_{2}+2\right) }{\alpha _{2}}\right] ^{\frac{1}{\alpha _{2}+2}}\left(
t-t_{0}\right) ^{\frac{1}{\alpha _{2}+2}}.
\end{equation}
The scale factor can be obtained from $da/d\phi=\left[(1/\sqrt{
6})\coth G\right] a=\left(\phi /2\alpha _{2}\right)a$, and it has exponential dependence
on the scalar field and time
\begin{eqnarray}
\nonumber
a=a_{0}\exp \left( \frac{\phi ^{2}}{4\alpha _{2}}\right) =\\
a_0\exp \left\{
\frac{1}{4\alpha _{2}}\left[ \frac{\left( \alpha _{2}+2\right) \sqrt{2V_{0}}%
}{\alpha _{2}}\right] ^{\frac{2}{\alpha _{2}+2}}\left( t-t_{0}\right) ^{%
\frac{2}{\alpha _{2}+2}}\right\} ,
\end{eqnarray}
were $a_{0}$ is an arbitrary constant of integration.

\section{The superpotential method}

The superpotential method for the standard
inflationary model (\ref{1.1}) was successfully applied for solving the SCEs (\ref{1.7})-(\ref{1.9}).
The main idea was \cite{Chervon:2000yg} to represent the SCEs in the form of the slow roll approximation. i.e., in the equations for a spatially-flat universe (we obtain them from (\ref{1.7})-(\ref{1.9}) by setting $\epsilon =0$)
\begin{eqnarray}\label{fried_H1-4}
H^2=\frac{1}{3}\left( \frac{1}{2}\dot{\phi}^2 +V(\phi)\right),\\
\label{fried_H2-4}
\dot{H}=- \frac{1}{2}\dot{\phi}^2, \\
\label{scal_field-4}
\ddot{\phi}+3H\dot{\phi}+V'(\phi)=0,
\end{eqnarray}
we should omit $\ddot{\phi},~~ \dot{\phi}^2$. After that, the SCEs in the slow roll regime take the following form
\begin{eqnarray}\label{fried_H1-4sr}
H^2 \simeq \frac{1}{3}V(\phi),\\
\label{fried_H2-4sr}
\dot{H}=- \frac{1}{2} \dot{\phi}^2, \\
\label{scal_field-4sr}
3H\dot{\phi}\simeq -V'(\phi).
\end{eqnarray}

To obtain the desired form of the equations, the potential of total energy \cite{Zhuravlev:1998ff}  (or the superpotential \cite{Chervon:1997yz}, \cite{Fomin:2017xlx}) was introduced,  as the sum of kinetic energy (in terms of scalar field argument $U(\phi)=\dot{\phi}$) and the potential energy
\begin{equation}
\label{def w}
W(\phi)=\frac{1}{2} U^2(\phi)+V(\phi),~~U(\phi) = \dot\phi .
\end{equation}
After this substitution, the SCEs take the following form:
\begin{eqnarray}
\label{ei-1w}
H^2=\frac{1}{3}W(\phi),\\
\label{ei-2w}
\dot{H}=-\frac{1}{2}U(\phi)^2,\\
\label{f-vw} 3H\dot\phi =-W'(\phi).
\end{eqnarray}

As we can see the system of equations above exactly reproduce the SCEs
in slow roll form (\ref{fried_H1-4sr})-(\ref{scal_field-4sr}) if we substitute $W(\phi)$ instead of $V(\phi)$ and $U(\phi)$ instead of $\dot{\phi}$.

As we know, any one from the presented exact SCEs in the slow roll form (\ref{ei-1w})-(\ref{f-vw}) can be derived as a consequence
(or a differential consequence) of the remaining two. In our
approach, we exclude from consideration  equation (\ref{ei-2w}) in
the first step.

Excluding from (\ref{ei-1w}) the Hubble parameter $H$, and inserting it
into (\ref{f-vw}), taking into account  the superpotential definition
(\ref{def w}), we have the consequence of (\ref{ei-1w}) and (\ref{f-vw})
in the form
\begin{equation}\label{u-w}
 \sqrt{3}U^2W^{1/2}=-W'.
\end{equation}
Integrating (\ref{u-w}) with respect to $W$, we obtain the relation:
\begin{equation}\label{w-int-u}
W=\frac{3}{4}\left( \int U(\phi)d\phi\right)^2,
\end{equation}
which leads to the new method of exact solution construction in
cosmology, viz., by suggesting that evolution of the scalar
field is given, one can determine the superpotential by solving
the integral on the right hand side of (\ref{w-int-u})
$ \int\dot\phi^2 dt $. Knowing $W$, one can find $H$ from the Friedmann
equation (\ref{ei-1w}) with the following relation
\begin{equation}\label{h-int}
H=\frac{1}{4}\sqrt{3}\left( \int U(\phi)d\phi\right).
\end{equation}
Then, by integration, one can find the scale factor $a(t)$.

Thus, the proposed method presents some combination of the two methods:
slow roll-like presentation of the exact equations \cite{Zhuravlev:1998ff} and
obtaining cosmological solutions for the given scalar field evolution
\cite{Barrow:1994nt}.

The advantage of the proposed method lies in the essential
simplification of the integration procedure: one needs to calculate
only one integral for obtaining a superpotential and Hubble
parameter. Then the potential $V$ (\ref{def w}), as well as the
scale factor $a(t)$, are calculated from related definitions. The
exhibition of the simplicity of the procedure,  and its effectiveness can be
found  in the applications of the method  to cosmology on the brane, in
phantom and tachyon fields \cite{Astashenok:2013lda},\cite{Chervon-Panina},\cite{Chervon-Panina-Sami}.
The two last  have a very restricted
number of exact solutions which can be essentially extended by
virtue of the superpotential method.

We can have a look at the superpotential method from another position.
The system   (\ref{fried_H1-4}) - (\ref{scal_field-4}) has three unknowns,
$\phi(t)$, $V(\phi)$ and $H(t)$ (or $a(t)$).  To solve it, one of these variables has to be
given a  priori.  It is customary to look for the solution for a given
$V(\phi)$, but as it is known, it is very difficult to solve the SCEs for a given potential exactly.

In the superpotential approach, it is proposed to reduce the equations to a simpler form 
which helps to solve them exactly.  In order to do this, let us consider the
superpotential function $W(\phi)$ defined in  (\ref{def w})
%
$$
W(\phi) =\frac{1}{2}U(\phi)^2+ V(\phi),~~U(\phi) = \dot\phi.
$$
%
Now, with the change of variable $dt\!=\!d\phi/\dot\phi$ and using the inverse  transformation from $U(\phi)$ to $\dot{\phi}$, one can obtain
\begin{equation}\label{v2}
\frac{d W}{d\phi}=\frac{d V}{d\phi} + \ddot\phi.
\end{equation}
Hence, the SCEs  (\ref{fried_H1-4}) - (\ref{scal_field-4}) can be rewritten in the slow roll form (\ref{fried_H1-4sr}) - (\ref{scal_field-4sr}).
Therefore we can try to solve SCEs in the superpotential presentation because the slow roll approximation was intensively studied.

The superpotential $W(\phi)$ ($>0$) shows
up as the main part of the potential function, driving the dynamics of the Hubble parameter $H$ or the scale factor.
To solve them, note that Eq. (\ref{ei-1w}) defines $\dot a/a$ as a
function of $\phi$, $H(\phi)$, which when inserted into Eq.(\ref{f-vw}),
gives the scalar field $\phi(t)$ as a function of $t$, at least in quadratures
\begin{equation}\label{cuadr}
- 3 H(\phi) \left( \frac{d W}{d\phi} \right)^{-1} d\phi =  dt.
\end{equation}

Finally, inserting $\phi(t)$ into eqs. (\ref{def w}) and (\ref{f-vw}) gives
$V(\phi)$ and $a(t)$, respectively, and the solution is completed.

Obviously,
one could simply have begun by giving $H$=$H(\phi)$, but it is usually desirable to
have some description of the potential instead, and for this reason it is
preferable to give $W(\phi)$.
One could also use $H(t)$ to determine $\phi(t)$, since
\begin{equation}
\frac{1}{2}\dot\phi^2 = -\frac{d H(t)}{dt}
\end{equation}
implies that
\begin{equation}
\Delta\phi(t) = \pm\int\sqrt{-2 \frac{d H(t)}{dt}} ~ {dt},
\end{equation}
where $\Delta\phi(t)=\phi-\phi_*,~~\phi_* $ is the integration constant.
So since $W=3H^2(t)$, a complete knowledge of $H(t)$ fully determines
the solution to the problem.

Also, we can obtain potential V from the superpotential W
\begin{equation}
V(\phi)=W(\phi)-\frac{W_{\phi}'^{2}(\phi)}{6W(\phi)}.
\end{equation}

\subsection{Examples of exact solutions}

Firstly, we consider the superpotential  $W=\lambda \phi^{2n}$.
For this type of superpotential, we have 
\begin{equation}\label{vlf2n-1}
\phi(t) =
\left[ \phi_0^{2-n} \pm
2n (n-2) \sqrt{\frac{\lambda}{3}} ~ (t-t_0) \right]^{\frac{-1}{n-2}},
\end{equation}
\begin{equation}\label{vlf2n-3}
V(\phi) = \lambda \phi^{2n} - \lambda \frac{2 n^2}{3}\phi^{2(n-1)},
\end{equation}
\begin{equation}\label{vlf2n-2}
a(t) = a_0 \exp\left\{ \frac{-1}{4n} \left[ \phi_0^{2-n} \pm
2n (n-2) \sqrt{\frac{\lambda}{3}} (t-t_0) \right]^{\frac{-2}{n-2}} \right\}.
\end{equation}

Another potential used in the literature is that of a hyperbolic cosine $W=V_0 \left( \cosh(\beta \phi)-1 \right)$
\cite{Matos:2009hf}.  The solution found there, with $\phi_0$=0, is
\begin{equation}\label{cosh-1}
\phi(t) = \frac{-2}{\beta} \mbox{arcsinh} \left( \tan \left(\sqrt{\frac{V_0}{6}} \beta^2 t \right) \right),
\end{equation}
\begin{equation}\label{cosh-3}
V(\phi) = V_0 \left( \cosh(\beta \phi)-1 \right) -
V_0 \frac{\beta^2}{6} \left( \cosh(\beta\phi) +1 \right),
\end{equation}
\begin{equation}\label{cosh-2}
a(t) = a_0 \cos^{\frac{2}{\beta^2}}\left( \sqrt{\frac{V_o}{6}}\beta^2 t\right).
\end{equation}

The expressions for generating functions by means of  superpotential $W$ are represented in the Table $1$.

\begin{widetext}
\vspace{4mm}

\begin{tabular}{|l|c|c|}
\hline
Number and authors & Kind of generating function & Connection with $W$ \\
\hline
$I\,\,\, \mbox{Ivanov (1981), Chervon, Fomin (2017)}$ & $H(\phi)=\frac{1}{\sqrt{3}}\left(F(\phi)+F_{\ast}\right)$ & $F(\phi)=\sqrt{W}-F_{\ast}$\\
\hline
$II \,\,\, \mbox{Chimento {\it et al},  (1993)}$ & $F(a)=V[\phi (a)]a^{6}$ & $F(a(\phi))=a^{6}\left[W-\frac{W_{\phi}'^{2}}{6W}\right]$ \\
\hline
$III \,\,\, \mbox{Schunck and Mielke, (1994)}$ &  $g(H)=V(H)-3H^2$ & $g(H(\phi))=-\frac{W_{\phi}'^{2}}{6W}$ \\
\hline
$IV \,\,\, \mbox{Kruger and Norbury, (2000)} $ & $F(\phi)=1+\left(\frac{\dot{\phi}^{2}}{V(\phi)}\right)$  & $F(\phi)=\frac{6W^{2}+W_{\phi}'^{2}}{6W^{2}-W_{\phi}'^{2}}$ \\
\hline
$V \,\,\, \mbox{Charters and Mimoso, (2010)}$ & $x(\phi)=\dot{\phi}/H$ & $x(\phi)=-\frac{W_{\phi}'}{W}$ \\
\hline
$VI \,\,\, \mbox{Harko {\it et al}, (2014)}$ & $\dot{\phi}=2V(\phi )\sinh ^{2}G(\phi )$ & $\coth G(\phi)=\frac{\sqrt{6}W}{W_{\phi}'}$ \\
\hline
\end{tabular}
\end{widetext}
\vspace{4mm}

\section{Conclusion}

The space of exact solutions in scalar field cosmology is very huge. In this work, we review exact solutions in inflationary cosmology and the method of finding such solutions. There are basically three methods of constructing such solutions. For any scale factor $a(t)$, we can find the potential and kinetic energy to be satisfied in the self-consistent system of Einstein and scalar field equations. Secondly, from the evolution of the scalar field
$\phi=\phi (t)$, we can find the scale factor $a(t)$ and the potential $V(t)$, thus  defining the exact solution. Thirdly, choosing the generation function, represented in this given review, one can, once again, obtain  a  great number of exact solutions.

Thus, as the next step of the investigation, we suggest  analysing the method of confrontation of theoretical  predictions from exact solutions with observational data.

\section*{Acknowledgements}

The authors are grateful to an anonymous referee for  qualitative comments and suggestions, which contributed to the improvement of this work.

I.V. Fomin was supported by RFBR grants 16-02-00488 A and 16-08-00618 A.


\end{document}